\documentstyle[preprint,aps,epsf]{revtex} 
\newcommand{\be}{\begin{equation}} 
\newcommand{\bef}{\begin{figure}} 
\newcommand{\eef}{\end{figure}}

\newcommand{\etal}{{\em et al. }} 
\newcommand{\ee}{\end{equation}} 
\def\spose#1{\hbox to 0pt{#1\hss}} 
\def\ltapprox{\mathrel{\spose{\lower  
3pt\hbox{$\mathchar"218$}} 
 \raise 2.0pt\hbox{$\mathchar"13C$}}} 
\def\gtapprox{\mathrel{\spose{\lower  
3pt\hbox{$\mathchar"218$}} 
 \raise 2.0pt\hbox{$\mathchar"13E$}}} 
\def\inapprox{\mathrel{\spose{\lower  
3pt\hbox{$\mathchar"218$}}     
 \raise 2.0pt\hbox{$\mathchar"232$}}} 
 
\begin{document} 
\title{On the fractal structure of galaxy 
 distribution and its implications for cosmology 
} 
\maketitle 
\centerline{} 
\centerline{\Large Yurij V. Baryshev } 
\centerline{Astronomical  
Institute of the Saint-Petersburg University, 198904, 
St.-Petersburg, Russia }  
\centerline{} 
 
\centerline{\Large Francesco Sylos Labini
\footnote{D\'ept.~de Physique Th\'eorique, Universit\'e de Gen\`eve,
24, Quai E. Ansermet, CH-1211 Gen\`eve, Switzerland}, 
Marco Montuori, Luciano Pietronero } 
\centerline{Dipartimento di Fisica, Universit\`a di Roma  
``La Sapienza'' P.le A. Moro 2, I-00185 Roma, Italy.} 
\centerline{INFM, Sezione di Roma 1} 
\centerline{} 
 
\centerline{\Large Pekka Teerikorpi} 
\centerline{ Tuorla Observatory,  
University of Turku, FIN-21500 Piikki$\ddot o$, Finland } 

  
\begin{abstract}  
Two fundamental empirical laws have been established in  
the analysis of 
galaxy space  distribution. 
First, recent analyses 
have revealed that the three dimensional distribution of 
galaxies and clusters is characterized by large scale
structures and huge voids: such a 
distribution shows
fractal correlations up  
to the limits of the available samples. 
This has confirmed the earlier de Vaucouleurs  
power-law density - distance 
relation, now corresponding  
to a fractal structure with dimension $D \approx 2$, at least, 
in the range of scales $ \sim 1 \div 200 \; Mpc$
($H_0 = 55 km/sec/Mpc$). 
An eventual cut-off towards homogenization has not been
yet identified.
Second, since Hubble's discovery, the linear redshift-distance law  
has been well established within $200 Mpc$ and also much deeper. 
The co-existence of these laws within the same scales is a challenge 
for the standard cosmology where the linear Hubble law is a strict 
consequence of homogeneity of the expanding universe. 
This puzzle is now sufficiently strong to raise doubts for the standard 
cosmology.  
\end{abstract} 
  
\newpage  
 
\section*{1. Introduction}

The basic assumption of the standard  
Big Bang cosmology is the Einstein's 
Cosmological Principle which, in fact, is 
the hypothesis that the universe is spatially homogeneous and isotropic 
on large scales (see e.g. Weinberg, 1972; Peebles, 1993). 
Homogeneity of  matter distribution plays the central role in  
the expanding universe model,  
because it  implies that the recession velocity is  
proportional to distance. 
This means that linear velocity-distance law  
$v_{exp}=Hl$,  
identified by the observed Hubble law, is valid at scales 
where matter distribution can be considered on average uniform.

For a long time, astronomers used 2-dimensional photographic plates 
of the sky as the basic means for the galaxy structures
studies 
without no direct observations of the 
3-dimensional large-scale matter  
distribution.  Then the linear Hubble law inferred from redshift 
measurements at various distances, was interpreted as indirect  
evidence for average homogeneity at the same scales (see Peebles, 1993).

Recently, several 3-dimensional maps of galaxy  
distribution  have become available, based on massive 
redshift measurements.
Surveys such as CfA, SSRS, Perseus-Pisces, IRAS, 
LEDA, APM-Stromlo, Las Campanas, and ESP for galaxies, and Abell and 
ACO for galaxy clusters
have detected remarkable  
structures such as filaments, sheets and voids.   
The galaxy maps now  
probe scales up to $200 Mpc$ and they show that the  
large-scale structures are common features of the local universe. 
Fig.\ref{fig1} illustrates this change from flat sky vault to 
the deep 3-dimensional space.

Pietronero and collaborators  (see the review by Sylos Labini \etal,1998
for a comprehensive discussion of the subject) by using the methods
of modern statistical physics, have shown that, in the above mentioned
surveys, 
galaxy distribution exhibits fractal behavior with dimension 
$D\approx 2$ 
at least up to $200 Mpc$. 
This implies that, in the context of the 
standard model, one would expect strong deflections from  
the linearity of the redshift-distance relation 
in the range of scale where 
the inhomogeneities are present.  However, observations show an almost 
strictly linear Hubble law at the same distance scales where fractal 
distribution has been established.  In this article we explore 
this unexpected  inconsistency of the standard model.

\section*{2. De Vaucouleurs Law and universal fractal structure}

In the history of astronomy, parallel to the idea of homogeneity which 
was adopted by Newton, the alternative view of a hierarchic structure 
of the Universe has existed since it was created by 
Swedenborg, Kant and Lambert in the 18th century. 
They imagined the majestic picture of 
the hierarchy of stars forming clusters forming  galaxies  
which in turn are clustered into yet larger systems and so on,  
either indefinitely (Swedenborg , Kant) or stopping at some level (Lambert). 
This idea was revived by Fournier 
d'Albe and Charlier who 
discussed a hierarchy where the mass within radius $r$ varies  
as $M \sim r$ (for historical reviews see Mandelbrot, 1977 
and Baryshev \& Teerikorpi, 1998a).

De Vaucouleurs (1970) studied the possibility, firstly mentioned  by 
Carpenter (1938) and Kiang (1967), that there is a  
universal density-radius power law as a basic factor in  
cosmology, reflecting a hierarchic distribution.  He suggested  
the law: 
\be 
\label{e7} 
\rho(r) = \rho_l \left(\frac{r}{r_l}\right)^{-\gamma} 
\ee 
where $\rho(r)$ is the mass density in the sphere of radius $r$, 
$\rho_l$ and $r_l$ are the density and radius at the lower cutoff 
of the hierarchy, and  
$\gamma$ is nowadays called the fractal co-dimension or correlation 
exponent.  This is related to fractal dimension $D$ as: 
\be 
\label{e8} 
\gamma = 3 - D \; .
\ee

The concept of fractal was introduced by Mandelbrot (1977).  
The mathematical framework of fractal geometry allows one to study 
in a quantitative way the properties of irregular systems. 
Fractal  
structures are characterized by clusters and voids at all scales,  
and they are intrinsically non-analytical: their scaling behavior is  
described by power law functions.  The density-radius relation 
(Eq.\ref{e7}) is an example of such a behavior. 
 
It is a remarkable property of fractal structure that the 
power law scaling for the density (Eq.\ref{e7})  
holds for any structure point.  This means that the average density
decreases 
not only away from us, but from any other galaxy. This ensures 
the statistical equivalence 
of all the observers in the distribution, which is required by any 
reasonable cosmological principle (see below).  

In a fractal structure the   conditional (average) density from an  
occupied  point  behaves as (Pietronero 1987) 
\be 
\label{l6} 
\Gamma (r) = S(r)^{-1}\frac{ dN(r)}{dr} = \frac{D}{4\pi } B r ^{-(3-D)} 
\ee 
where $\:S(r)$ is the area of  a spherical shell of radius $\:r$,
thickness $dr$ 
and   $\:dN(r)/dr$ is the number of points in the shell
and $B$ is constant related to the lower cut-off (see Eq.\ref{e7}).
 
 If the distribution is fractal up to a certain distance $\lambda_0$, 
and then it becomes homogeneous, we have that: 
$$ 
\Gamma(r) = \frac{BD}{4 \pi} r^{D-3} \; \;\; r < \lambda_0 
$$ 
\be 
\label{e327b}  
\Gamma(r)= \frac{BD}{4 \pi} \lambda_0^{D-3} = const. \; \;\;r \geq \lambda_0 
\ee

From Eq.\ref{e327b} it results that the conditional density
is the appropriate statistical tool able to study
scale invariant systems as well as homogenous ones.
On the other hand, the {\it standard correlations analysis}
is strongly based on the assumption that 
the system has reached homogeneity in the sample
analyzed. In fact, the standard 
approach (Peebles, 1990; Davis \& Peebles, 1983)
identifies 
a length-scale, called the {\it "correlation length"} 
$\:r_{0}$,  and defined by the relation: 
\be 
\label{e330} 
\xi(r_{0})= 1 
\ee 
where 
\be 
\label{e331} 
\xi(r) = \frac{<n(\vec{r_{*}})n(\vec{r_{*}}+ \vec{r})>}{<n>^{2}}-1 
\ee 
is the (standard) 
two point correlation function. 
(In Eq.\ref{e331} $n(\vec{r})$ is the local number density).
The basic point that    Pietronero and collaborators 
(Pietronero, 1987; Coleman \& Pietronero, 1992; 
Sylos Labini, \etal 1998)  stressed, 
is that the mean density, $<n>$, 
used in the normalization of $\:\xi(r)$ (Eq.\ref{e331}), 
is not a well defined quantity 
in the case 
of self-similar distribution and it is a direct function of the sample size. 
Hence only in the case that 
the homogeneity  has been reached well within the sample 
limits, the $\:\xi(r)$-analysis is meaningful, otherwise 
the a priori assumption of homogeneity is incorrect and the 
characteristic lengths, like $\:r_{0}$, became spurious. 
 
For example  
the expression of the $\:\xi(r)$ in the case of 
fractal distributions is 
\be 
\label{e332} 
\xi(r) = \left(\frac{3-\gamma}{3}\right) 
\left( \frac{r}{R_{s}} \right)^{-\gamma} -1 
\ee 
where $\:R_{s}$ is the depth of 
the spherical volume where
average density is computed. 
From Eq.\ref{e332} it follows that:
 
i.) the so-called correlation 
length $\:r_{0}$ (defined as $\:\xi(r_{0}) = 1$) 
is a linear function of the sample size $\:R_{s}$:
\be 
\label{e333} 
r_{0} = \left(\frac{3-\gamma}{6}\right)^{\frac{1}{\gamma}}R_{s} 
\ee 
and hence it is a spurious quantity without  physical meaning:  it is 
simply related to the sample finite size.

ii.) $\:\xi(r)$ is power law only for: 
\be 
\label{e334} 
\left(\frac{3-\gamma}{3}\right) 
\left(\frac{r}{R_{s}}\right)^{-\gamma}  \gg 1 
\ee 
hence for $r \ll r_0$: for larger distances there is a clear deviation 
from the power law behavior due to the definition of $\xi(r)$. 
This deviation, however, is just due to the size of 
the observational sample and does not correspond to any real change 
of the correlation properties. It is clear that if one estimates the 
exponent of $\xi(r)$ at distances $r \ltapprox r_0$, one 
systematically obtains a higher value of the correlation exponent 
due to the break of $\xi(r)$ in the log-log plot 
(see for more details Sylos Labini \etal, 1998).

The analysis 
performed by $\xi(r)$ is therefore mathematically inconsistent, if 
a clear cut-off towards homogeneity has not been reached, because 
it gives an information that is not related to the real physical 
features of the distribution in the sample, but to the size of the 
sample itself. 
\bigskip


The new approach proposed by
Pietronero and collaborators 
has been applied to the above mentioned redshift surveys, 
with the result that fractal structures extend to scales larger than  
$\sim 200 Mpc$.  In particular, the correlation function  
has been found to have a power law behavior up to 
distance $\sim 300 Mpc$, 
establishing that Eq.\ref{e7} 
holds  with  
$\gamma=1 \pm 0.2$, or in terms of fractal dimension  
$D =2 \pm 0.2$. 
The eventual cut-off towards homogeneity
(the length scale $\lambda_0$ in Eq.\ref{e327b})
has not been yet identified. The long-range 
correlations detected are limited only by the
size of the available catalogs.
The results from the $\Gamma(r)$ function analysis in the 
presently available redshift surveys are given in 
Tab.\ref{table1} (from Pietronero \etal, 1997).

Usually analysis of the space distribution is based on 
distances estimated from redshifts which always include a non-distance 
part due to peculiar velocities of galaxies.  Recently, Teerikorpi 
\etal (1998) have derived 
the radial galaxy spatial distribution around our Galaxy, utilizing 
over 5000 Tully-Fisher photometric distances from the KLUN program 
(KLUN = Kinematics of the Local Universe).  First results give evidence 
for a decrease in the average density consistent 
with the fractal dimension $D \approx 2.2$ 
in the distance range 
$20 \div  200 Mpc$.  
This may be regarded as a new, independent confirmation 
of the results of Sylos Labini \etal (1998).

\section*{3. Hubble Law} 
 
In his classic paper, Hubble (1929) found a roughly linear 
relation between the spectral line displacement  
$z =(\lambda_{obs}-\lambda_{em})/\lambda_{em}$ 
of the line emitted by a far away galaxy $  \lambda_{em}$,
and its distance $r$.  The empirical 
Hubble Law may be written as 
\be 
\label{e5} 
cz = H_0 r \; ,  
\ee 
where $c$ is the velocity of light and $H_0$ is the Hubble 
constant.  As an observationally established relation, the Hubble 
law does not refer to any interpretation of redshift, though Hubble  
had in mind especially the de Sitter effect (Smith, 1979). This  
effect in the static space, as well as the  
space expansion and Doppler mechanisms for redshift, yields 
at first order to Eq.\ref{e5}.  If redshift is interpreted as a motion 
effect, then 
\be 
\label{e6} 
V \approx H_0 r 
\ee 
where $V$ is either space expansion velocity $v_{exp}$  
or ordinary velocity of a body moving 
in the Euclidean space.  Usually this 
velocity-distance relation is called the Hubble Law, but it is 
more correct to regard it as 
the redshift-distance relation of Eq.\ref{e5} 
(Harrison, 1993).  This is based on the primarily measured quantities 
(redshift and distance), while velocity is inferred from redshift 
in the frame of some cosmological model.

Since its discovery, the validity of the Hubble law has been  
confirmed in an ever increasing distance  
interval where local and more remote distance indicators may be 
tied together.  Recently, several new distances have been measured  
to local galaxies using observations of Cepheid variable stars, 
thanks to the Hubble Space Telescope programmes (Freedman, 1996; 
Tammann, \etal 1996).  
Along with previous Earth-based Cepheid distances,  
methods like Supernovae Ia and Tully-Fisher have been better 
calibrated than before and confirm the linearity with good 
accuracy up to $z \approx  0.1$.  Brightest cluster galaxies 
trace the Hubble law even deeper, up to $z \approx 1$, and 
radio galaxies have provided such evidence at still larger 
redshifts (Sandage, 1995).

It is well known that there are small deviations $\delta V$ from 
the Hubble velocity $V_H$, connected with local mass 
concentrations such as 
the Virgo Cluster, and, possibly the Great Attractor. 
However, these perturbations are still only of the order 
$\delta V/V_H \sim 0.1$, while in the general  
field the Hubble law has been suggested to be quite smooth, 
with $\delta V$ around $50 km/s$ (Sandage, 1995; Karachentsev \&  
Makarov, 1996).

The observed Hubble law is derived for our Galaxy as the center, and 
it does not necessarily imply that it is observed similarly from 
other galaxies.  However, if our Galaxy is not in a privileged position 
in the Universe, then one must conclude that the Hubble law is 
universal.  The small scatter even at the small distances of a few 
Mpc also seems to be a general feature: the structure and kinematics 
of our Local Group is typical of several other small groups (Karachentsev, 
1996; Governato \etal, 1997).

There have been proponents of global non-linearity of the  
Hubble law, which is expected in some non-standard cosmological 
models (e.g. Segal, 1976).  However, it has been convincingly shown 
that such deviations from linearity are easily produced by 
a statistical distortion effect known as the Malmquist bias in  
magnitude limited samples.  When one by-passes this problem using 
suitable methods, the linearity is recovered (Teerikorpi, 1997). 
The second main advancement is concerned with measurement of the value 
of the Hubble constant $H_0$. Since Hubble's $559 \; km/s/Mpc$,  
$H_0$ has decreased and according to the most 
recent studies (Tammann \etal, 1996; Theureau \etal, 1997), 
seems to have stabilized around $55 \; km/s/Mpc$.  In this article, we 
shall use this value for distance estimates from redshift.

Without actual knowledge of the matter distribution, the linearity 
and small scatter of the observed  
Hubble law for field galaxies would make one easily guess 
that the galaxies are uniformly distributed: as it was asserted 
above, this is the basis for the linear Hubble law in the standard 
cosmology.  In fact, it  
has been a common supposition that when the Hubble law was found  
in the nearby space, one finally had entered a cosmologically  
representative region of the Universe. At the same time, 
it has been clear  
that at small distances where Hubble found his  
relation, the galaxy distribution is quite inhomogeneous. 
Though, it has been believed that beyond some, not too large 
distance, the distribution should become uniform.

\section*{4. The Hubble - de Vaucouleurs Diagram}

As we have already mentioned,  
studies of the 3-dimensional galaxy universe have shown 
that de Vaucouleurs' prescient view on the matter distribution 
is valid at least in the range of scales  $\sim 1 \div 200 \;Mpc$. 
The Hubble and de Vaucouleurs laws describe very different  
aspects of the Universe, but both have in common universality and  
observer independence.   
This makes them fundamental cosmological laws and it is important 
to investigate the consequences of their coexistence at 
similar length-scales.  In Fig.\ref{fig2} we display these laws together.

A representative Hubble law has been taken from Fig.4 of Teerikorpi 
(1997), based on Cepheid distances to local galaxies, Tully-Fisher 
distances from the KLUN programme, and Supernovae Ia distances. 
The behavior of the conditional density 
(De Vaucouleurs law) presented in Fig.\ref{fig2} was taken from  
Sylos Labini \etal (1998).

The puzzling conclusion from Fig.\ref{fig2} is that the strictly linear  
redshift-distance relation is observed deep inside the  
fractal structure.  This empirical fact presents a profound 
challenge to the standard model where the  
homogeneity is the basic explanation of the Hubble law, and "the  
connection between homogeneity and Hubble's law was the first success of  
the expanding world model" (Peebles \etal, 1991).  
This also reminds us 
the natural reaction  of several authors: 
"In fact, we would not expect any neat relation of 
proportionality between velocity and distance  
[for such close galaxies]" (Weinberg, 1977).

However, contrary to the expectations, modern data show a good linear  
Hubble law even for nearby galaxies.  How unexpected this actually is, 
can be expressed quantitatively for the standard model and is discussed 
below.

\section*{5. Implications for Friedmann models} 
 
\subsection*{5.1. Homogeneity and isotropy as the basis for Hubble law} 
 
According to the standard Big Bang model the universe obeys to the 
 Einstein's 
Cosmological Principle: it is homogeneous, isotropic 
and expanding (Weinberg, 1972; Peebles, 1988; 1993).  
Homogeneity of matter distribution is the  
central hypothesis 
of the standard cosmology because it allows one to introduce the space  
of uniform curvature in the form of the Robertson-Walker line element 
\be 
\label{e1} 
ds^2 = c^2 dt^2 - R(t)^2 
[dy^2 + f(y)^2 (d \Theta^2 + \sin^2 \Theta d \Phi^2)] \; .
\ee 
Here $y, \Theta, \Phi$ are  
comoving space coordinates, $t$  
is comoving cosmic time, 
$f(y) = \sin(y), y, \sinh(y)$  
for curvature-constant values $k = 1, 0, -1$  
respectively, $R(t)$ is the scale factor, and $c$ is the velocity of light. 
This line element leads immediately to a linear  
relation between velocity and proper distance (Robertson, 1955).  
Indeed, consider a comoving body at a fixed coordinate distance 
from a comoving observer.  At cosmic $t$,
let $l= R(t) y$ be the  proper distance 
from the observer.
The expansion velocity $v_{exp} = dl/dt$, 
defined as the rate of change of the proper distance $l$,
is 
\be 
\label{e2} 
v_{exp} = H l = c \cdot  \frac{l}{l_{H}}
\ee 
where $H = \dot{R}/R$ is the Hubble constant and $l_H = c/H$ 
is the Hubble  
distance.  In this way, the linear velocity-distance relation of  
Eq.\ref{e2}  
is an exact formula for all Friedmann models and a  rigorous consequence 
of spatial homogeneity. 
In particular, for $l > l_H$, the expansion velocity $v_{exp} > c$.   
Such an apparent violation of  
special relativity is consistent with general relativity 
(Harrison, 1993).

In the expanding space the wavelength of an emitted photon is  
progressively stretched, so that the observed redshift $z$ is  
given by Lemaitre's redshift law 
\be 
\label{e3} 
z = \frac{R(t_{obs})}{R(t_{em})} - 1 
\ee 
which is a consequence of the radial null-geodesic of the line element  
Eq.\ref{e1}. For $z \ll 1$ Eq.\ref{e3} yields  
$z \approx dR/R \approx H_0 dt \approx l/l_H$,  
and from Eq.\ref{e2} one gets the approximate velocity-redshift  
relation that is valid for small redshifts 
\be 
\label{e4} 
v_{exp} \approx cz \; . 
\ee

The relations between $v_{exp}, l$, and $z$  
have been recently discussed in 
an especially clear manner by Harrison (1993) who emphasized that the  
expansion velocity-redshift relation differs from the  
relativistic Doppler 
effect.  So, the space expansion redshift 
mechanism in the standard model is quite distinct from the usual  
Doppler mechanism.  We stress this points, because in the literature 
these two redshift mechanisms are often confused.

In the context of the standard  
cosmology, it has been natural to interpret the Hubble Law  
as a reflection of Eq.\ref{e6} and Eq.\ref{e4}  
and to regard the coefficient of  
proportionality $H_0$ in Eq.\ref{e5}  
as the present value of the theoretical Hubble constant $H$  
from Eq.\ref{e2}.

The behavior of the scale factor $R(t)$ is governed by the  
Einstein equations which describe how the geometry of space 
changes under the influence of a mass distribution.  In the case 
of homogeneity and zero-pressure, they lead to the Friedmann 
Equation: 
\be 
\label{ynew1} 
\frac{d^2R} {{dt}^2}  = - \frac{4\pi}{3} G R \rho 
\end{equation}
This equation gives two basic parameters of the Friedmann models: 
the Hubble parameter $H(t) = [dR(t)/dt] [1/R(t)]$,  
the value of which at the 
present time $t_0$ is equal to the Hubble constant $H_0$, and the 
density parameter $\Omega = \rho / \rho_{crit}$ where the critical 
density is defined as $\rho_{crit} = 3 H^2 / 8\pi G$.  
 
Note that the Friedmann 
Equation coincides exactly with the equation for the Newtonian force 
acting on the unit mass on the surface of a spherical ball with uniform 
density $\rho(t)$ and radius $R(t)$.  Hence, there is a tight analogy 
between the behavior of the scale factor in Friedmann models and 
the expanding radius of the finite Newtonian ball.  This analogy was first 
noticed by Milne (1934) and later discussed by Layzer (1954) and McCrea 
 (1955). 
 
 \subsection*{5.2. Gravitational growth of density fluctuations}


In order to give  a rough estimation of the gravitational
growth of density fluctuations and accompanying deflection from
the linear velocity-distance relation, one can utilize the
linear perturbation (LP) approximation (see Peebles, 1980; 1993).

We consider now the case of an expanding universe,
where an average density is well defined and has a constant value $\rho_0$.
In such a case, by neglecting relativistic 
effects  and the terms depending on pressure, 
according to the linear approximation, there is a velocity
deflection $\delta V$ from the unperturbed Hubble flow $V_H=H_o r$
in the scale where the density perturbation is $\delta \rho$.
In the case of zero cosmological constant
and {\it spherical mass distribution}, this deflection has grown during
the Hubble time to the present value which is (Eq.20.55 from
Peebles, 1993):
\be
\label{pek1}
\frac {\delta V} {V_H} = \frac { 1}{ 3} {\Omega_0}^{0.6}
\frac {\delta \rho}   {\rho_0}
\ee
where $\Omega_0 = \rho_0/\rho_{crit}$ is the density parameter of
the Friedmann model.
This approximation holds in the limit $\delta \rho/\rho_0\ll 1$.


Let us consider a two-component model for the density distribution in
a Friedmann universe.  
First, there is the component which exhibits fractal behavior up to  a maximum 
scale, and which we call
$\lambda_0$.  At larger scales   
this component is homogeneous with an average density 
$\rho_{lum}$.
The second  component is  dark matter, homogenous at all scales, 
with density
$\rho_{dark}$.
For such a model there is a definite constant density at all scales larger
than $\lambda_0$.  This density $\rho_0$ is the sum of $\rho_{lum}$ and
$\rho_{dark}$.  This means that the behavior of this model at scales
larger than $\lambda_0$ is identical to that of the Friedmann model for which
\be
\label{yn1}
\Omega_0 = \Omega_{dark} + \Omega_{lum} \; .
\ee
At such large scales the Hubble law is unperturbed.  
The density distribution of luminous matter   for
scales $r < \lambda_0$, can be written as 
\be
\label{yn2}
\rho_{lum} (r) =   \rho_{lum} \left( \frac{\lambda_0 }{r}\right)^{\gamma}
\ee
where $\gamma=3-D$ as usual.
For the scales $r \ge \lambda_0$ we have that
\be
\label{yn3}
\rho_{lum} (r)  =  \rho_{lum}= \rho_{lum} (\lambda_0) \;.
\ee

The density contrast can be written as 
\be
\label{yn4}
\frac{\delta \rho  }{\rho_0} = \frac{ \rho_{lum} (r) + \rho_{dark} - \rho_0}
{\rho_0}
\ee
In terms of the Friedmann density parameters as defined above, this becomes,
at scales smaller than $\lambda_0$:
\be
\label{pek2}
\frac{\delta \rho }{  \rho_0} = 
\frac{\Omega_{lum} }{\Omega_0} 
\left(  
\left(\frac{r}{\lambda_0} \right)^{-\gamma}
- 1  \right) \; .
\ee
At scales larger than $\lambda_0$ the density contrast 
clearly vanishes. A useful distance scale can be defined by the relation
\be
\label{f1}
\frac{\delta \rho(\lambda_{0.5}) }{  \rho_0} = 0.5 \;.
\ee
At scale larger than $\lambda_{0.5}$ we have that 
$\delta \rho / \rho_0 \ll 1 $ and Eq.\ref{pek1}
holds. From Eq.\ref{f1} we easily obtain
\be
\label{f2}
\lambda_{0.5}= \lambda_0 \left( 1 + \frac{1}{2} \frac{\Omega_0}{\Omega_{lum}}
\right)^{-\frac{1}{\gamma}}\;.
\ee

By using the linear approximation (Eq.\ref{pek1}), we may obtain
a rough estimation of the expected deflection from the Hubble law
in the two component model, for the scale
in which the density contrast is less than 1. 
Although the linear approximation
is valid only for $\delta \rho/\rho_0 \ltapprox 1$,
the obtained results give a first quantitative indication
of the effects of self-similar fluctuations.
Moreover the assumption of spherical 
mass distribution is a rough one, and it holds only 
for average quantities. In the case 
of real fractals deviation from spherical symmetry
can play an important role, at least at small scale,
(Sylos Labini \etal, 1998c).
Under these approximations,  
the radial velocity measured by an average observer
at scale $r < \lambda_0$ is (from Eq.\ref{pek1})
\be
\label{pek3}
V_{obs} = V_H \left(1-\left(\frac{1}{3}\right)
{\Omega_0}^{-0.4} 
 \Omega_{lum}
 \left( 
 \left( \frac{r}{\lambda_0} \right)^{-\gamma}
 - 1 \right)\right) 
\ee
Actually, this is the prediction averaged over many observers in different
fractal structure points (galaxies).  For any particular observer, there
will be a deflection from this average law.

We take the maximum scale $\lambda_0$ of
fractality and the fractal dimension $D$ from the observed
cosmological de Vaucouleurs Law and calculate the expected deflections
from the Hubble law in our two-components Friedmann model.  

In Fig.3 we show three theoretical predictions for the velocity
deflection in the case where
the observed fractal structure contains all the matter, i.e. when
$\Omega_{lum} = \Omega_0$.  We have fixed $\lambda_0 = 200 Mpc$ and
fractal dimension $D = 2$. 
In this case the linear approximation (from Eq.\ref{f2}) holds for
$r \gtapprox  130 Mpc$. At smaller scales we should
consider non-linear effects which are not simple to be treated.
However we should expect even stronger velocity
perturbations, due to the highly inhomogeneous 
structures distribution.
The predictions correspond to three values
of the cosmological density parameter $\Omega_0 = 1, 0.1, 0.01$.  From
Fig.3 it follows that such Friedmann models, purely fractal within
$200 Mpc$, are excluded if $\Omega_0 \gtapprox 0.01$.  This confirms the
previous suggestions that small $\Omega_0$ is needed for hierarchic
models (Wertz, 1971; Haggerty \& Wertz, 1972, and Sandage, \etal 1972 - hereafter STH).
Fang \etal (1991) used a similar two-components model as a possible
explanation for claimed differences in the values of the Hubble constant
as derived from near and distant galaxies.  Because new data confirm
the linearity of the Hubble law, implications of the two-component model
are different.

For instance, there is a possibility to save the Friedmann universe with the
critical density parameter $\Omega_0 = 1$.  It was implied
already by STH that dark matter, uniformly filling the
whole universe and decreasing the relative density fluctuations, could
reconcile the observed fractal structure with the linear Hubble law.  
However, they did not give a quantitative estimate of the amount of dark
matter needed.  With
the new data on the Hubble and de Vaucouleurs laws, we can
derive the lower limit for the amount of the needed uniform dark matter.
In Eq.\ref{pek3} we fix 
$\Omega_0 = 1$ and
let $\Omega_{lum}$ have different values.  Fig.4 gives the developed
version of the STH test, now showing that $\Omega_{dark}$ should be  larger than
0.99.  If the actual maximum scale of fractality is larger than $200 Mpc$
(with $D=2$),
then the amount of luminous matter may be in conflict with the Big Bang
nucleosynthesis prediction for baryonic matter.  For example, if
 $\lambda_0 \gtapprox 1000 Mpc$ 
(as suggested by Sylos Labini \etal, 1998)
then
$\Omega_{lum}$ will be probably less than 0.001.

It should be emphasized that this estimate of the amount of dark matter
is independent on the physics of the early universe.  It also does not
depend on the determination of mass-to-luminosity ratio of galaxies.

\section*{6. Discussion} 

The surprising coexistence of the Hubble and de Vaucouleurs laws  
contains many still unexplored cosmological implications. Above 
we have discussed its importance from the viewpoint of the standard model 
where the Hubble law is more fundamental and requires homogeneity. 
In this case, there must be a homogeneously 
distributed dark matter, producing the linear Hubble law, and 
the observed fractal structure is confined to the sparse 
luminous matter having its properties 
(fractal dimension and maximum scale) due to some initial conditions.

It is also possible that the HdeV-diagram gives rise to a whole new 
direction in cosmological research, based on the more general  
Conditional Cosmological 
Principle (see below).  In this case, the de Vaucouleurs law is as  
fundamental as the Hubble law and encompasses all matter.  Then the Hubble 
law is not related to homogeneity, rather to isotropy.  A new theoretical 
problem appears concerning the physical reason for the observationally 
revealed universal value of $D \approx 2$.  We point out that the old  
question on the gravitational part of the cosmological redshift needs to 
be reanalyzed in the context of fractal structures.


Isotropic fractals allow the universe to be isotropic and 
inhomogeneous, without sacrificing 
the equivalence of all the observes. In this 
case it is possible to formulate a weaker version of
the Cosmological Principle, called 
the Conditional Cosmological Principle  (Mandelbrot, 1977; 1998; 
Coleman \& Pietronero, 1992). 
This implies an asimmetry between space points occupied by the 
structure and empty points. In fact, a fractal structure is
statistically the same from every point of the system.
Such a property correspond to Local Isotropy rather than
to homogeneity.


 Could it be an accident that the observed fractal dimension for 
galaxy distribution is so close to 2?  
Perhaps there is some physical 
 significance in the $D \approx 2$ fractal structure. 

\subsection*{6.1. Gravitational part of cosmological redshift}

A fully self-consistent dynamics of a fractal cosmology has not 
yet been developed.  However, 
one can at least suggest that a Hubble law due to expansion is not expected, 
because of the global non-uniformity.  Hence, if the linear Hubble 
 law is, however, observed, its origin may be dominated by 
 something other than expansion.

Gravitation is the only known mechanism, other than Doppler 
effect and space expansion, to produce the redshift of galaxy  spectra. 
Bondi (1947) showed that the cosmological redshift may be divided 
in two parts, due respectively to relative motion and to 
gravitation effect. 
He emphasized that the spectral shift depends not only on the motion 
of source and observer, but also on the global matter distribution around 
the source, extending up to the observer.  The gravitational part of 
the cosmological redshift was also considered by Zeldovich and Novikov  
(1984).  They asserted that the gravitational shift in the homogeneous 
universe should increase as the square of distance, hence it does not 
obey the linear Hubble law. Furthermore, they concluded that it is a blueshift 
and not a redshift.  This followed from their choice of a coordinate system, 
with the origin at the observer and the light sources rested on the surface 
of a sphere around the observer.

However, this choice is not unique. 
If one puts the origin on the source and the observer is on the surface of 
sphere surrounding the source, one similarly derives redshift.  In favor 
of the last choice, one can present a causality argument: the emission of  
a photon by the source precedes its absorption by observer's light detector.  
After time $\tau$ from the emission of a photon, all observers at the 
surface of the sphere with radius $c\tau$, around the source, have a 
probability to detect the photon.  Hence the shift will be redshift. 
The sign of the gravitational shift may be also extracted from the 
general Mattig's relation between proper distance and cosmological 
redshift, and it corresponds to redshift (Baryshev \etal, 1994).

It has been pointed out by Baryshev (1981) that 
inside a hierarchical structure with fractal dimension $D = 2$,  
the gravitational part of the cosmological redshift  
is linearly proportional to distance, producing a linear $(z_{gr} - r)$ 
relation.  Indeed, if $\rho(r) \sim 1/r$,  
then the mass is spherical sample of radius $r$ scales as $M \sim r^2$,  
while the  gravitational potential  behaves as 
$\phi(r) \sim M/r \sim r$.  
So the gravitational redshift is $z_{gr} \sim \phi/c^2 \sim r$. 
In this case the gravitational fractal contribution to the value 
of the Hubble constant is: 
\be 
\label{ee8} 
H_{grav} = 2 \pi G \rho_l r_l/c 
\ee 
where $\rho_l$ and $r_l$ are the density and radius of the lower  
cut-off in the fractal structure 
(see Eq.\ref{e7}), $G$ is the gravitational constant, 
and $c$ is the velocity of light.

The gravitational part may be 
unobservable small or a significant fraction of the observed Hubble 
constant, depending on the product $\rho_l \cdot  r_l$ which characterizes 
the fractal structure. 
As a convenient reference value, one can use 
 $\rho_l r_l = 1/2 \pi g cm^{-2}$, 
in which case $H_{grav} =  G/c = 69 km/s/Mpc$. 
For an average galaxy with 
$\rho_l = 5 \cdot 10^{-24} gcm^{-3}$ and $r_l=10 kpc$ 
this condition is fulfilled.

It should be noted that the cosmological tests proposed for 
determining the nature of the observed redshift, such as 
independence from wavelength, Tolman surface brightness test, 
time delay, blackbody temperature (see Sandage, 1995), cannot 
distinguish
between expansion and gravitational redshifts  
(Baryshev \etal, 1994). 
The astrophysical 
possibilities to detect this gravitational redshift  
will be discussed elsewhere by Baryshev \& Teerikorpi(1998b). 
 
\subsection*{6.2 Theoretical approaches to $D \approx 2$}
 
We briefly summarize various theoretical approaches which
have considered the fractal properties of matter distribution.
 
An interesting 
analysis leading to $D \approx 2$ was made by Schulman and Seiden 
 (1986).  They proposed a model of galaxy formation.  In their model the 
 original process of galaxy formation occurs in analogy to the process of 
 propagating star formation within our Galaxy.  The birth of one galaxy 
 stimulates the birth of nearby galaxies from the primeval cosmic gas.  
 If this process occurs near its percolation threshold then a hierarchical 
 structure is created.  The critical fractal dimension for this process is 
 quite close to 2 (about 1.95).

It is also interesting to note that within the so-called plasma cosmology 
originated by H. Alfven there is a possibility to understand the 
existence of hierarchic structures starting from certain general principles. 
Here the fractal dimension 2 appears as a consequence of electromagnetic 
processes in plasma (Lerner 1986).


  Recently, in a paper by De Vega \etal (1996a) 
(see also De Vega \etal, 1996b; 1998) 
it was shown that 
gravitating 
 mass points in statistical equilibrium are distributed with fractal 
 dimension close to 2. Note that the energy density of the gravitational 
 field (which is 
positive and localizable in the field theory - Baryshev 1996) will be homogeneous for 
fractal structures with $D = 2$.  This may be a way to understand  
the above result on the equilibrium state of self-gravitating point 
mass distribution having $D\approx 2$. 

\section*{7. Conclusions}

Investigation of the large scale distribution of galaxies in 
the universe is now in a new phase, which is characterized 
by new observational data and new methods of analysis.  It has 
become an especially hot and debated topic in cosmology, because 
the revealed fractality contradicts Cosmological Principle 
in the sense of homogeneity. 
 Above we have discussed the fractality and its 
implications for cosmology.  Our main conclusions are: 
 
\begin{itemize}

\item Observations show that there is a fractal distribution of 
galaxies, having fractal dimension $D\approx 2$ in the scale 
range  from $1 Mpc$ to, at least, $200 Mpc$.  While there is a general
agreement on the small scale fractal properties of galaxy distribution,
the actual value of $D$ and 
the eventual presence of an upper cut-off, are still matter of 
debate
(see e.g. Pietronero \etal, 1997; Davis, 1997; Guzzo, 1997; 
Coles, 1998; Sylos Labini \etal, 1998a; 1998b).
 
\item The traditional statistical analysis 
based on the assumption of homogeneity
(i.e. $\xi(r)$), 
should be replaced by the more general methods
of modern statistical physics.
Such methods are able to characterize scale-invariant
distributions
as well as regular ones.
 
\item An isotropic fractal distribution
is fully compatible with the reasonable
requirement of the equivalence of all the observers.
Hence the Standard Cosmological Principle,
which requires isotropy {\it and} homogeneity,
may be replaced by the  Conditional  Cosmological Principle.
In such a case the condition of local isotropy around
any structure point, without the assumption of 
analyticity of matter distribution, does not 
imply the homogeneity of matter distribution.

 
\item The paradox of linear Hubble law within the fractal de Vaucouleurs 
density-distance law is sharpened with the new data: strong deflections 
from the Hubble flow are expected in the framework of the standard 
Friedmann model. 
 
\item From a developed version of the old Sandage-Tammann-Hardy test we 
derive the minimum amount of the uniform dark matter, $\Omega_{dark} = 0.99$, 
which is consistent with the presently known Hubble and de Vaucouleurs 
laws.  This result is independent of the early universe physics.  If the 
maximum scale of fractality is larger than $200 Mpc$, this test may be 
regarded as crucial for the standard cosmology. 
 
\item It is noted that within fractal structure there is a net 
 gravitational redshift which follows a linear 
redshift - distance law for fractal dimension 2.  The significance 
of its contribution to the cosmological redshift depends on 
the amount of mass coupled to fractals and requires further study.


\end{itemize}

\bigskip 
 
{\bf Acknowledgements} 
 
Yu.B. thanks for support by the Russian program "Integration", project 
N. 578.  This work has been supported by Academy of Finland (project 
"Cosmology in the Local Universe"). 
F.S.L. M.M.  and L.P. thank  
L. Amendola, H. Di Nella 
and A. Gabrielli for useful collaborations. 
F.S.L. is grateful to M. Joyce and 
 R. Durrer for useful discussions and suggestions. 

\def\rapj{Ap. J.} \def\rmnras{MNRAS} 
\def\raa{A\& A}

\newpage 
 
\begin{table} 
\centering 
\begin{minipage}{140mm} 
\caption{\label{table1} 
The volume limited samples, constructed from redshift surveys 
and needed in the analysis, are characterized by the following 
parameters: 
- $R_s (Mpc)$ is the depth of the catalogue 
- $\Omega$ is the solid angle 
- $R_1 (Mpc)$ is the radius of the largest sphere 
that can be contained in the catalogue volume. 
This gives the limit of statistical validity of the sample. 
- $r_0(Mpc)$  is the length at which $\xi(r) \equiv 1$. 
- $D$ is the fractal dimension of the sample. 
The value of $r_0$ refers to the deepest subsample. 
The CfA2 and SSRS2 data are not yet available  
(distances are expressed in $Mpc$ and $H_0 = 55 km/sec/Mpc$). 
} 
\begin{tabular}{|c|c|c|c|c|c|}  
     &      &  &    &              &     \\ 
\rm{Sample} & $\Omega$ ($sr$) & $R_s  $ & $R_1  $ 
& $r_0  $ & $D$ \\ 
\hline
    &       &    &    &    		   &       \\ 
CfA1 & 1.83  & 160 & 40 & 12  & $1.9 \pm 0.2$    \\ 
CfA2 & 1.23  & 260& 60 & 20 & $\approx 2.0$               \\ 
PP   & 0.9   & 260& 60 & 20 & $2.0 \pm 0.1$  \\ 
SSRS1& 1.75  & 240& 70 & 24 & $2.0 \pm 0.1$      \\ 
SSRS2& 1.13  & 300& 100 & 30 & $\approx 2.0$               \\ 
Stromlo-APM& 1.3  & 200& 60 & 20 & $2.2 \pm 0.1$      \\ 
LEDA & $4 \pi$ & 600& 300& 90 & $2.1 \pm 0.2$   \\ 
LCRS & 0.12  & 1000& 36 & 12  & $1.8 \pm 0.2$    \\ 
IRAS $1.2 Jy$ & $4 \pi$ & 160 & 80 & 9& $2.0 \pm 0.1$\\ 
ESP  & 0.006 & 1400& 20 & 10  & $1.9 \pm 0.2$  \\ 
     &       &               &      &  &       \\  
\hline
\end{tabular} 
\end{minipage} 
\end{table} 

\newpage 
 
\bef  
\epsfxsize 10cm 
\centerline{\epsfbox{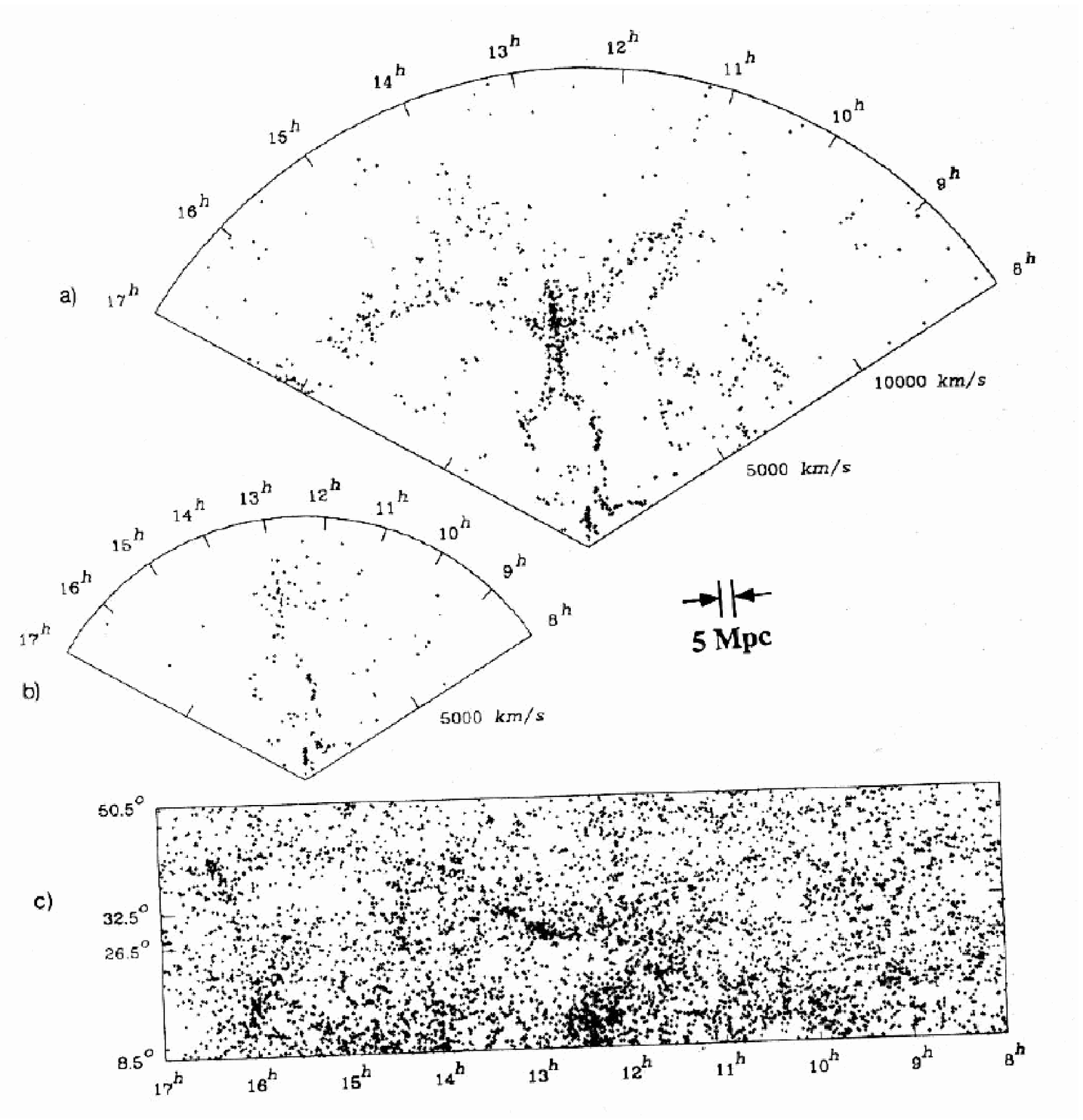}} 
\caption{\label{fig1} 
A slice of  the three dimensional galaxy distribution (old CfA1 catalog  
{\it (a)} 
and new CfA catalog {\it (b)}) 
compared with the corresponding {\it (c)} 
angular distribution (the portion between  
$26.5^{\circ}$ and $32.5^{\circ}$) - from the de Lapparent \etal 1988. 
Note that the angular distribution appears relatively  
homogeneous while the real distribution in space is much more irregular.  
In particular this picture points out the so-called Great Wall  
which extends over the entire sample (at least $340 Mpc$). 
We also show the size of the galaxy "correlation-length" ($r_0 = 10 Mpc$) 
derived from the standard analysis (here distances correspond to the 
value of the Hubble constant $H_0 = 55 km/sec/Mpc$). The more  
general analysis that we discuss in the text implies instead that  
an eventual correlation length should be larger than any observable 
structure, i.e. $>  340 Mpc$ and that the present data show well defined 
fractal properties up to the sample limits (from De Lapparent \etal, 1998). 
} 
\eef

\bef 
\epsfxsize 6 cm 
\centerline{\epsfbox{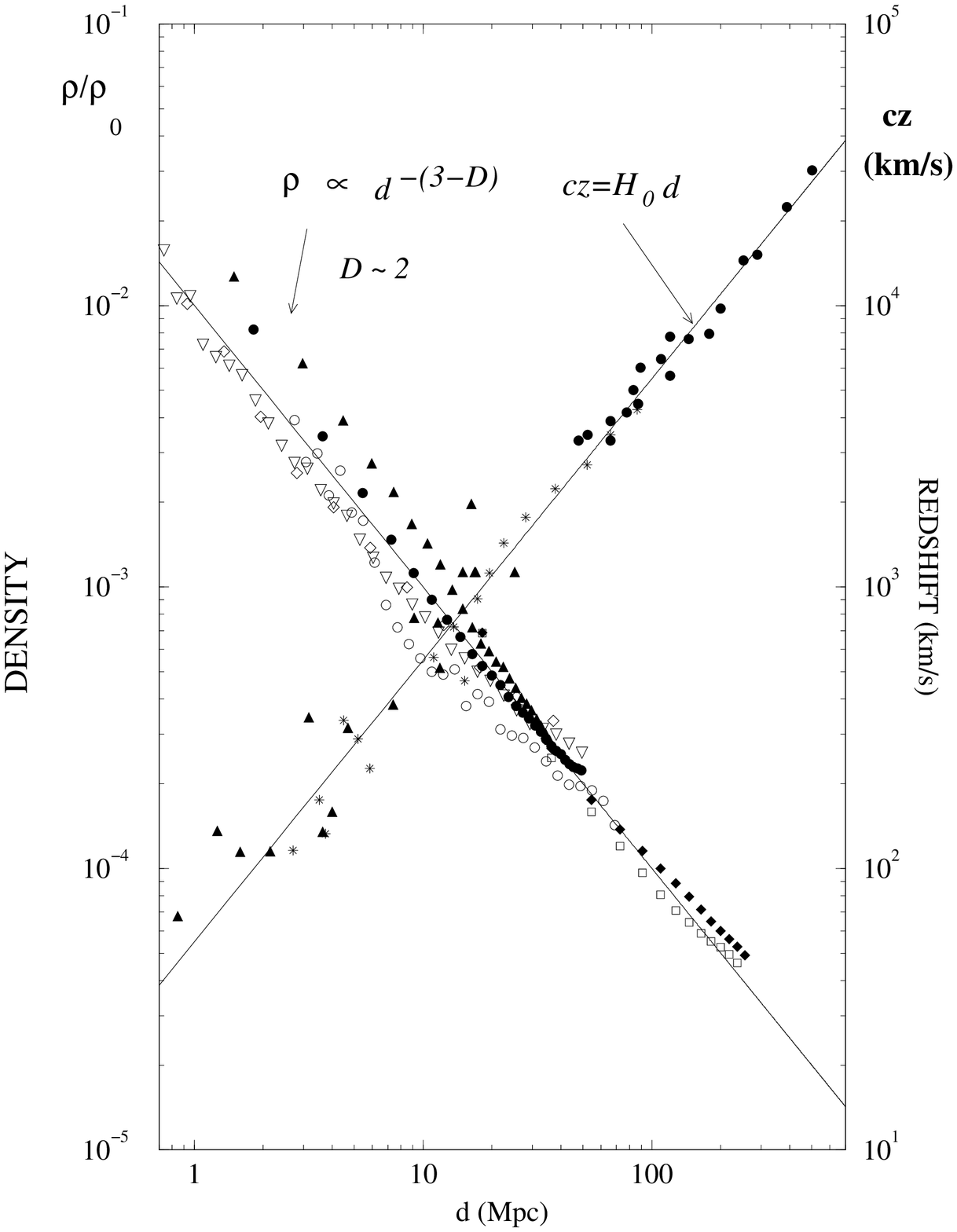}} 
\caption{\label{fig2} 
 Hubble redshift-distance and de Vaucouleurs density-distance  
laws in the distance scales from $1$ to  
$500 Mpc$ 
 ($H_0=55 km/s/Mpc$).  
 The Hubble law (increasing from left to right) is constructed 
from: galaxies with Cepheid-distances for $cz > 0$ (triangles),  
galaxies with Tully-Fisher 
(B-magnitude) distances (stars), galaxies with SNIa-distances for $cz > 3000 
km/s$ (filled circles). 
 TF-distance points are generally averages of a few tens of galaxies 
from the "unbiased plateau" of the method of normalized distances.  
Redshift $cz$ is reduced to the Local Group center and contains 
the small correction due to the Virgo infall velocity field.  The solid 
line  corresponds to the Hubble law with  $H_o = 55 km/s/Mpc$. 
The de Vaucouleurs law (decreasing from left to right) in the normalized 
form is constructed from  
the computation of the conditional average density in the following 
 redshift surveys: 
CfA1 (crosses), Perseus-Pisces, LCRS (filled diamonds), 
ESP (triangles left) 
and LEDA (Sylos Labini \etal, 1998). 
The normalization between the different densities 
takes into account the different magnitude limited of the various  
redshift surveys. 
The dotted line  corresponds to the de Vaucouleurs law with 
 correlation exponent $\gamma = 1$, i.e. $D=2$. } 
\eef 
  
\bef 
\epsfxsize 12cm 
\centerline{\epsfbox{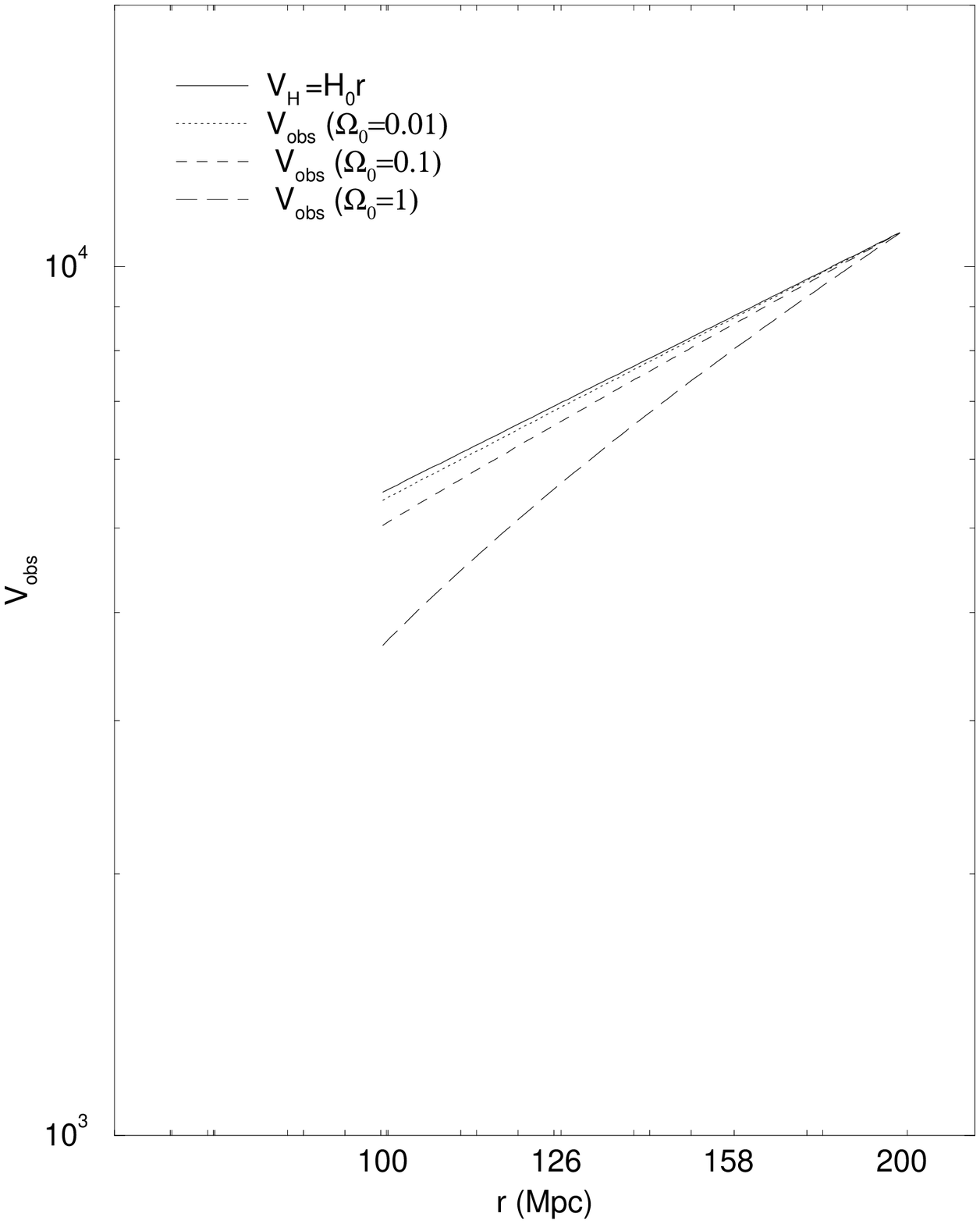}} 
\caption{\label{fig3}
We show three theoretical predictions for the velocity
deflection in the case where
the observed fractal structure contains all the matter, i.e. when
$\Omega_{lum} = \Omega_0$.  We have fixed $\lambda_0  = 200 Mpc$ and
fractal dimension $D = 2$.  The predictions correspond to three values
of the cosmological density parameter $\Omega_0 = 1, 0.1, 0.01$.
} 
\eef 

\bef 
\epsfxsize 12cm 
\centerline{\epsfbox{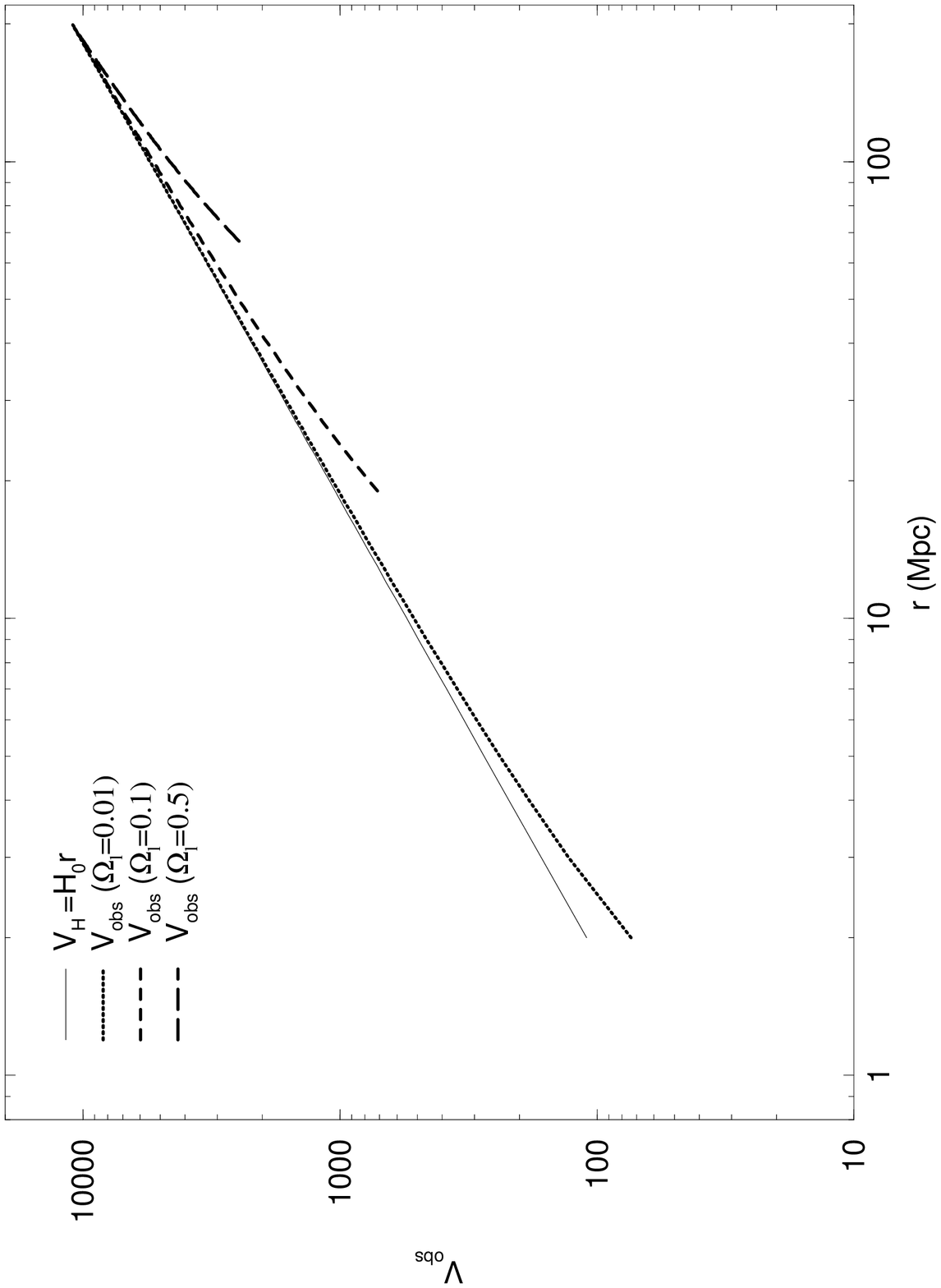}} 
\caption{\label{fig4} In this case the total density is $\Omega_0=1$.
We show the behaviour of $V_{obs}$ derived by the 
linear perturbation  approximation in the case 
$\delta \rho / \rho_0 \ll 1$, for various values of $\Omega_{lum}$.
The fractal dimension of luminous matter is $D=$ up to
$\lambda_0 = 200 Mpc$. } 
\eef 
\end{document}